\documentstyle[12pt]{article}

\renewcommand{\a}{\alpha}

\renewcommand{\d}{\delta}

\newcommand{\ep}{\epsilon}
\newcommand{\g}{\gamma}

\newcommand{\si}{\sigma}

\newcommand{\p}{\partial}

\newcommand{\non}{\nonumber}

\newcommand{\anpj}[1]{{ \it Ann.~Phys.~(N.Y.)}~{\bf {#1}}}

\newcommand{\plbj}[1]{{ \it Phys.~Lett.}~{\bf {#1B}}}
\newcommand{\npbj}[1]{{ \it Nucl.~Phys.}~{\bf B{#1}}}

\begin{document}

\newcommand{\beeq}{\begin{equation}}
\newcommand{\eneq}{\end{equation}}
\newcommand{\bear}{\begin{eqnarray}}
\newcommand{\enar}{\end{eqnarray}}
\newcommand{\xd}{\p_t {\vec X}}
\newcommand{\fdt}{\rm "field\; dependent\; term"}

\begin{titlepage}
\begin{flushright}
\parbox{1.in}
{
 IFT-20/92\\
 October 1992}
\end{flushright}
\vspace*{.5in}
\begin{centering}
{\Large \bf Quark Spin and the $\Theta$-term \\
for the QCD String  }
\footnote{Work supported, in part,
by Polish Government Research Grant KBN 2 0165 91 01.}\\
\vspace{2cm}
{\large        Jacek Pawe\l czyk}\\
\vspace{.5cm}
        {\sl Institute of Theoretical Physics, University of Warsaw,\\
        Ho\.{z}a 69, PL-00-681 Warsaw, Poland.}\\
\vspace{.5in}
\end{centering}
\begin{abstract}
We describe a way in which spin of quarks can
enter a consistent QCD string theory.
We show that the spin factor of the 4d massless, spin 1/2
fermions is related to the self-intersection  number of a
2d surfaces immersed in the 4d space. We argue that the latter quantity
should appear in a consistent description of the QCD string. We also
calculate the chiral anomaly and show that
the self-intersection number corresponds to
the topological charge $F{\tilde F}$ of QCD.
\end{abstract}
\end{titlepage}

Since the formulation of the quantum mechanics in terms of path integrals
\footnote{There exists extensive literature to the
subject contained e.g. in [1],[2],[3]}
there  exists a challenging problem of description of the particle spin
within the formalism.
 The recent papers on the subjest [2,3] elaborate on the
so-called spin factors which were introduced by Polyakov [4].
The subject is interesting by its own but also it can shed more light on the
possibility of inclusion of similar factors in the string theory.
Here we want to focus on the relation between the spin factor of the
4d massless, spin 1/2 fermion and the self-intersection number of a 2d
surface (with boundary) immersed in the 4d (euclidean) space-time.
The self-intersection number, denoted by $I$, appeared in the context
of the rigid string.
In [5] it was proposed to consider 2d
dynamics of the rigid string with so-called $\theta$-term, $2\pi i \theta$.
The hope was that for $\theta=\pi$ this term will help to
preserve the rigidity as relevant degree of freedom of the low energy
vibrations of the string. There were also suggestions that
the $\theta$-term corresponds to QCD $\theta$-term [6,7].

In this paper we shall discuss the way in which spin of quarks can
enter a consistent QCD string theory.
We show that after proper identification the  spin factor
of the massless 4d fermion equals to $e^{2\pi i\,I}$, where
$I$ is self-intersection number of a
2d surface $M$ immersed in 4d space. The fermion lives on the boundary
$\p M$. This suggests that
inclusion of the $\theta$-term at $\theta=\pi$ is equivalent to
taking into account 4d spin quantum numbers. In other words one can say
that the boundary of the string become 4d spin 1/2 fermion. This is
very desirable property because we believe that quarks live on the
boundary of the QCD string. We also introduce new anticommuting
variables on the boundary $\p M$ and derive the result for the
chiral anomaly. This result was obtained in [6] in a different, not purely
stringy way.

We shall consider the first quantized
massless spinning particle with spin $1/2$ moving in
d-dimensional space. The relevant action has the following form
[8,4]:
\beeq
S=\frac{1}{2} \int_0^1 dt\, \left( \frac{1}{e}(\xd)^2 -
\frac{1}{e^2}{\vec \psi} {\p_t {\vec \psi}}
+\frac{1}{e^2} \chi {\vec \psi}\xd \right)
\eneq
where $e$ denotes 1d graviton field, $\chi$ anticommuting gravitino,
${\vec X}$ position of the particle and ${\vec \psi} $ is
the supersymmetric partner of ${\vec X}$.
The crucial role is played by  the supersymmetry transformations
of the fields:
\bear
\d {\vec X}&= & \a{\vec \psi}/e \non\\
\d{\vec \psi}& = & \a(\xd - \frac{1}{2e}\chi{\vec \psi}) \non\\
\d\chi& = & -2{\p_t \a}\\
\d e& = & -\a\chi .\non
\enar
With help of the transformations (2) we shall show that one of the components
of the field ${\vec \psi}$ can be gauged away. Moreover, the gravitino
will work as
a Lagrange multiplier reducing the number of anticommuting fields ${\vec \psi}$
by one. Integration over the remaining anticommuting degrees of freedom
 leads to the spin
factors. Altogether we shall see that the description of the d-dimensional
massless spinor leads to the same spin factor as for the massive
(d-1)-dimensional spinor [2,3,4].

The Minkowski space
is the most natural frame  for the description of massless particles
with spin in the standard quantum field theory approach. However, here
we want to concentrate on closed paths thus we choose to work in
the euclidean space.
At each point of the path there is a natural
decomposition of the tangent space to
the space-time into 1d space spanned by $\xd$ and $d-1$ dimensional
space spanned by vectors ${\vec r}(t),{\vec n}^i(t)$,  $i=1,...d-2$.
The basis of the linearly independent vectors
$\{\xd,{\vec r},{\vec n}^i\}$ respects
\bear
\xd\, {\vec r} & {\not =} &0\non\\
\xd\, {\vec n}^i & = & {\vec r}\, {\vec n}^i=0\\
{\vec n}^i\, {\vec n}^j & = & \d^{ij}\non
\enar
The only restriction on the
{$\xd,{\vec r}$} sector of the basis is the
requirement that these vectors should be linearly independent and that
$\xd {\vec r}\,{\not =} 0$  everywhere. One can take
for convenience e.g. $\xd {\vec r}=1$.

The anticommuting field ${\vec \psi}$ can be decomposed in this basis in
the  following manner.
\beeq
{\vec \psi}=\phi_0\xd+\phi_r {\vec r} +\sum_{i=1}^{n-2} \phi^i {\vec n}^i
\eneq
With the help of the supersymmetry transformations (2)
we can immediately get the transformation rules for the components
$\phi^i$. From (2) and (3) we get
\bear
\d{\vec \psi}& =& \d\phi_0\,\xd+\phi_0\,\d\xd+\d\phi_r\, {\vec r} +\phi_r\,\xd+
\d\phi^i\, {\vec n}^i+\phi^i\,\d {\vec n}^i\non\\
& =& (\d\phi_0+...)\xd+(\d\phi_r+...){\vec r}+(\d\phi^i+...){\vec n}^i
\enar
where dots denote field dependent terms.
Comparison with (2) plus linear independence of the basis vectors
yields
\beeq
\d \phi_0=\a + \fdt.
\eneq
The above equation means that the filed $\phi_0$ can be completely gauged
away. The Fadeev-Popov determinant is trivial in this case.

Now we go to the discussion of the gravitino term. In the new basis it has
the following form:
\beeq
\frac{\chi}{e^2}{\vec \psi}\xd=\frac{\chi}{e^2}(\phi_0\xd^2+
\phi_r({\vec r}\xd))
\eneq
After the gauge fixing $\phi_0=0$, this term reduces to
$\chi\phi_r\xd {\vec r}/e^2$.
After redefinition of the gravitino $\chi\to e^2\chi/(\xd {\vec r})$
we simply get $\chi\phi_r$. The crucial point here is that $\xd
{\vec r}{\not =}0$.
The functional integration over the gravitino introduces
functional Dirac delta function $\d(\phi_r)$ under the path integral
, which effectively puts $\phi_r=0$ everywhere.
In this way the kinetic term of the fermionic part of the path integral
takes the form:
\beeq
{\vec \psi}{\p_t {\vec \psi}}=\phi^i{\p_t \phi}^i +C^{ij}\phi^i\phi^j
\eneq
where sum over $i,j=1,...d-2$ is understand and
$C^{ij}={\vec n}^i {\p_t {\vec n}}^j$.

Before we do the functional integral over the fermions $\phi$ we have
to change the functional measure: $D{\vec \psi}=D\phi\,J$.
The jacobian (J) of the transformation is
$det^{-1/2}(g^{\mu\nu}\p_t+C^{\mu\nu})$, where
$g^{\mu\nu}={\vec n}^\mu {\vec n}^\nu$ is the metric in
the new coordinate system, $C^{\mu\nu}={\vec n}^\mu {\p_t {\vec n}}^\nu$
and now ${\vec n}^{\mu}$ denotes all the basis vectors (3), $\mu=1\dots d$.
The jacobian is trivial because $C^{\mu\nu}$ is a pure gauge field.

Now we can perform functional integral over $\phi^i,\; i=1,...d-2$.
We consider a closed path with $\phi$'s respecting antiperiodic
boundary conditions (AP).
\beeq
\int_{AP}D\phi^i \; e^{1/2\int (\phi^i{\p_t \phi}^i +C^{ij}\phi^i\phi^j)}
=Tr\left\{ T\;{\rm exp}\left(   -\frac{1}{8} \int C^{ij}[\g^i,\g^j]
\right)\right\}
\eneq
where $T$ denotes the path ordered product and $\g's$ are Dirac matrices.
We also redefined $\phi^i\to e\phi^i$.
This expression represents so-called spin factor of the path integral
for the massless particle with spin 1/2. The spin factor is the trace
over the SO(d-2) group element, which is known to be relevant for the
classification of massless spin states.

{}From now on we shall consider only the 4d space-time.
In the case d=4 we can take the following Dirac matrices of interest:
$\g^1=1\otimes\si^1,\;\g^2=1\otimes\si^2$, so $[\g^i,\g^j]=
2i\ep^{ij}1\otimes\si^3$.
The spin factor is now\footnote{The sign ambiguity of the spin factor
[2] is resolved when we consider 2d surfaces for which the given
closed path will be a boundary}:
\beeq
Tr\left\{{\rm exp}\left(-\frac{i}{4}\int\;C^{ij}\ep^{ij}\;1\otimes \si^3
\right)\right\}
\eneq
We suppress the path ordering $T$ because the spin factor (10) is abelian.
It is the $4\times 4$ matrix which elements equals
${\rm exp}\{\pm i/4\int {\vec n}^i {\p_t {\vec n}}^{~j}\ep^{ij}\}$.
The integral in the exponential factor measures the rotation
angle (twist) of the frame $\{{\vec n}^1,{\vec n}^2\}$.
In the case of a massive particle the factor under the trace of (10)
belongs to a non-abelian group, which for the 4d spacetime is SO(3) [2].

Now we want to show that the above spin factor can be represented as
a surface functional integral. It is known that for 2d surfaces
immersed in the 4d space there exists a topological invariant counting
self-intersections of the given surface. The invariant is given by the formula
[5,6]
\beeq
I=-\frac{1}{16\pi}\int_M \,d^2\xi\sqrt{g}g^{ab}\p_a t_{\mu\nu}
\p_b {\tilde t}^{\mu\nu}
\eneq
where ${\vec X}={\vec X}(\xi^a)$ defines the immersion,
$g_{ab}$ is 2d induced metric on the surface $M$,
$g_{ab}=\p_a {\vec X}\p_b {\vec X}$ and
$t_{\mu\nu}=\ep^{ab}\;\p_a {\vec X}\p_b {\vec X}/\sqrt{g}$.
The formula for the self-intersection number $I$ can be put into different
, more suitable for our reasoning, form. We introduce gauge fields of
the vectors normal to the surface ${\vec N}^i$, (i=1,2), defined by
\beeq
A_a^{ij}(\xi)={\vec N}^i(\xi)\p_a {\vec N}^i(\xi).
\eneq
We also define their field strength tensor:
$F_{ab}^{ij}=\p_a {\vec N}^i \p_b {\vec N}^j
-\p_b {\vec N}^i \p_a {\vec N}^j$. Geometrically,
$F_{ab}^{ij}$ define an element of the second cohomology class $H^2(M)$.
Now the intersection number is
\beeq
I=\frac{1}{16\pi}\int_M\, d^2\xi\,\ep^{ab}\ep^{ij}F_{ab}^{ij}
\eneq

If the surface $M$ has a boundary $\p M$,
thus $H^2(M)=0$, the expression for $I$ simplifies to
\beeq
I=\frac{1}{8\pi}\int_{\p M}\,dt\, \ep^{ij} \,{\vec N}^i\p_t {\vec N}^j
\eneq
and now $I$ is no more a topological invariant.
Thus, $I$ is just the sum over the twists of the normals of all
the boundaries.
In the work [5] Polyakov proposed to add to the
rigid string action an $\theta$-term which is $2i\theta I$, i.e. he
proposed to consider
\beeq
\int D{\vec X}\;e^{-S[X]}\;e^{2i\theta I}
\eneq
{}From what we
said it is clear that for the open strings (i.e. surfaces with boundary)
at $\theta=\pi$ quantity ${\rm exp}\{2\pi iI\}$  can be identified
with the spin factor of the boundary (10), after identification of
$C^{ij}$ with $A_a^{ij}\xi^a$.
In this way we arrive at the interesting interpretation of the
self-intersection number:
if the surface is bosonic the $2i\theta I$ term at $\theta=\pi$
makes its boundary fermionic in the 4d sense. This suggests that
for $\theta=\pi$ $\theta$-term can simulate quarks at
the ends of the QCD string.
At $\theta=0$ the boundary term does not
give any contribution i.e. it stays to be bosonic.

All what was said till now can be summarized by the formula
\beeq
\int D{\vec X}\;e^{-S[X]}\;
Tr\left\{{\rm exp}\left(-\frac{i\theta}{4\pi}
\int_{\p M}dt\;A_a^{ij}\p_t \xi^a(t)\ep^{ij}  \right)\right\}
\eneq
where the trace under the functional integral is interpreted as
the spin factor of a massless particle (quark, if $\theta=\pi$)
living on the boundary $\p M$ of the surface $M$.
The functions $X^\mu$ define immersion of the surface $M$
into 4d euclidean space.
The field $A_a^{ij}(t)$ is a natural gauge field (of normals)
which appear on the string world-sheet $M$.

Now we introduce new anticommuting variables on the boundary
of the string. We denote them by $\Psi^a$, a=1,2.
We also introduce a new interaction of the fermionic path discussed above
with the string world-sheet.
The proposed modification arises naturally as
an interaction of $\Psi$'s with the external gauge field ( here it is
the gauge field of normals $A_a^{ij}$) if one treats them as superpartners of
2d coordinates $\xi^a$. In this way we get 1d locally supersymmetric
action for the interaction with $A_a^{ij}$ [8].
Thus the action for the boundary is
\beeq
S_{\p M}=-\frac{1}{2} \int_0^1 dt\left[\frac{1}{e}
g_{ab}\Psi^a{\p_t \Psi}^b+
\frac{i\theta}{2\pi}\ep^{ij}\left( A_a^{ij} {\p_t \xi}^a+
\frac{1}{2}F_{ab}^{ij}\Psi^a\Psi^b\right)\right]
\eneq
where $e$ is 1d metric of the boundary $\p M$. The action is 1d
reparametrization invariant so we can fix the symmetry setting $e=L$
(L is the moduli parameter of the boundary).

In a string theory with the boundary contribution (17) one can find out the
chiral anomaly. We shall consider $M$ with single boundary i.e.
$\p M$ is connected manifold.
The anomaly is given by the trace of the $\g_5$ with evolution
operator in the limit $L\to 0$ [10].
In the first quantized language it corresponds to evaluation of the
functional integral with 1d fermions respecting periodic (P) boundary
conditions in the limit $L\to 0$.

We shall not do the bosonic part
of the the integral concentrating on the fermionic contribution.
The anomaly comes from the zero modes $\Psi_0^a$ (i.e. constant
fields) of
$\Psi^a$ which appear for the periodic (P) boundary conditions.
The expression  for the anomaly reads:
\beeq
\lim_{L\to 0} Tr \left[\int d\Psi_0^1 d\Psi_0^2 \int_0^1 dt
\frac{i\theta}{8\pi}\ep^{ij}F_{ab}^{ij}(\xi_0)\Psi_0^a\Psi_0^b\right]
\int D\Psi' e^{-S'_{\p M}}
\eneq
where "prime" denotes omission of the zero modes, $\xi_0$ is the position
of the loop in the limit $L\to 0$. The only nontrivial contribution
from $S'_{\p M}$ which survive this limit is the topological term
corresponding to the spin factor. Finally
we get that the anomaly is proportional to
\beeq
\frac{\theta}{\pi}\frac{ N_f}{N_c}\ep^{ij}\ep^{ab}
 F_{ab}^{ij}\,{\rm exp}\left\{-\frac{i\theta}{8\pi}\int_M d^2\xi
F_{ab}^{ij}\ep^{ij}\ep^{ab}\right\}
\eneq
where $\p M=0$.
In the above formula the factor $N_f$ is the number of
different fermion species of $\Psi$'s
\footnote{One can discuss many-flavour case introducing
another type of anticommuting variables [10,11] or considering
matrix valued action for the boundary.}.
 The factor $1/N_c$  was introduced by hand due to the arguments from
large $N_c$ expansion which says that amplitudes with fermion loops
are suppressed by $1/N_c$ [12].

Chiral anomaly gives change of the functional integral
under infinitesimal (chiral) change of the phases of the spacetime
fermion fields. We see that this is equivalent to the change of the
$\theta$ in eq.(15) (but with $\p M=0$).
\beeq
\theta\to \theta(1+const.\frac{N_f}{N_c})
\eneq
For $\theta=0$ the anomaly vanishes. This suggests that in this case we
deal with a bosonic boundary.
The formula (20) differs from the standard QCD formula where we have
$\theta_{QCD}\to \theta_{QCD}+const.\;N_f/N_c$. This result holds
for $\theta_{QCD}$ close to zero, which we want to identify with
$\theta$ close to $\pi$. In this region one can take
$\theta=\pi-\theta_{QCD}$ and now both formulas agrees for small
enough $\theta_{QCD}$.

The same result can be obtained if one does not integrate over
$\phi$'s in (9). In this case the action is\footnote{The fields $\Psi^a$
were multiplied here by $\sqrt{e}$}

\beeq
S_{\p M}=-\frac{1}{2} \int_0^1 dt\left[
(g_{ab}\Psi^a{\p_t \Psi}^b+\phi^i\p_t\phi^i)-
\frac{\theta}{\pi}\phi^i\phi^j\left( A_a^{ij} {\p_t \xi}^a+
\frac{e}{2}F_{ab}^{ij}\Psi^a\Psi^b\right)\right]
\eneq
In order to calculate the anomaly the fields $\phi^i$ should also obey periodic
boundary conditions (P).  This procedure emphasize the 4d origin of the
anomaly with zero modes of $\phi$'s and $\Psi$'s playing the role
of $\gamma_5$.

We conclude the paper with several remarks and
speculations concerning application of the results to the QCD string.
First of all, it seems that the QCD string may correspond to
the rigid string with the $\theta$-term at $\theta=\pi$.
This is good news because there are some arguments which say that
one needs such a term in order to assure that the rigidity
is relevant for low energy vibrations of the string [5].
One might speculate that rigid string at $\theta=0$ may describe
the pure QCD or QCD coupled to scalar bosons.
In this paper we also add new anticommuting variables.
After coupling to
the gauge field of normals they reproduce expression for
the chiral anomaly with the self-intersection number playing the
role of topological charge $F{\tilde F}$ of the QCD.
\vskip1cm
{\bf Acknowledgement}

I want to thank my colleagues R.Budzy\'nski, Z.Lalak and M.Spali\'nski
for their kind interest in this work.
\newpage
\begin{center}
{\bf References}
\end{center}
\noindent
\begin{enumerate}
\item{M.Henneaux and C.Teitelboim, {\it in} Quantum Mechanics of
Fundamental Systems 2, eds. C.Teitelboim, J.Zanelli, Plenum Press,
New York,1988.}
\item J.Grundberg, T.H.Hansson and A.Karlhede, \npbj{347} (1990) 420.
\item{ M.A.Nowak, M.Rho and I.Zahed, \plbj{254} (1991) 94;
T.Jaroszewicz and P.Kurzepa, \anpj{210} (1991) 255;
I.A.Korchemskaya and G.P.Korchemsky, \plbj{257} (1991) 125.}
\item{ A.M.Polyakov, Gauge Fields and Strings, Harwood (1987);
 A.M.Polyakov, {\em in} Fields, Strings and Critical Phenomena,
Proc.Les Houches Summer School, Vol. IL, 1988, ed. E.Brezin and
J.Zinn-Justin, North-Holland, Amsterdam, 1990.}
\item  A.M.Polyakov, \npbj{268} (1986) 406.
\item P.O.Mazur and V.P.Nair, \npbj{ 284} (1986) 146.
\item A.P.Balachandran, F.Lizzi and G.Sparano, \npbj{263} (1986) 608.
\item {A.Barducci, R.Casalbuoni and L.Lusanna, {\it Nuova Cimento}
{\bf A35} 91976) 377; L.Brink, S.Deser, B.Zumino, P.Di Vecchia
and P.Howe, \plbj{64} (1976) 435; P.A.Collins and R.W.Tucker,
\npbj{121} (1977) 307; F.A.Berezin and M.S.Marinov, \anpj{104}
(1977) 307.}
\item J.\L opusza\'{n}ski, Rachunek spinor\.ow, PWN, Warsaw, 1985.
\item L.Alvarez-Gaume and E.Witten, \npbj{234} (1984) 269.
\item N.Marcus and A.Sagnotti, \plbj{188} (1987) 58.
\item G.'t Hooft, \npbj{72} (1974) 461; \npbj{75} (1974) 461.
\end{enumerate}

\end{document}